\documentclass[10pt,twocolumn,letterpaper]{article}

\usepackage{iccv}
\usepackage{times}
\usepackage{epsfig}
\usepackage{graphicx}
\usepackage{amsmath}
\usepackage{amssymb}
\usepackage{breqn}
\usepackage{adjustbox}
\usepackage[accsupp]{axessibility} 

\usepackage[pagebackref=true,breaklinks=true,letterpaper=true,colorlinks,bookmarks=false]{hyperref}

 \iccvfinalcopy 

\def\iccvPaperID{24} 

\ificcvfinal\fi

\begin{document}


\title{Generalized Zero Shot Learning For Medical Image Classification}


\author{Dwarikanath Mahapatra$^{1}$  \\
$^1$Inception Institute of Artificial Intelligence, UAE \\
{\tt\small dwarikanath.mahapatra@inceptioniai.org}
}


\maketitle

\begin{abstract}
In many real world medical image classification settings we do not have access to samples of all possible disease classes, while a robust system is expected to give high performance in recognizing novel test data. We propose a generalized zero shot learning (GZSL) method that uses self supervised learning (SSL) for: 1) selecting anchor vectors of different disease classes; and 2) training a feature generator. Our approach does not require class attribute vectors which are available for natural images but not for medical images. SSL ensures that the anchor vectors are representative of each class. SSL is also used to generate synthetic features of unseen classes. Using a simpler architecture, our method matches a state of the art SSL based GZSL method for natural images and outperforms all methods for medical images. Our method is adaptable enough to accommodate class attribute vectors when they are available for natural images.
\end{abstract}


\section{Introduction}
\label{sec:intro}

Medical image classification is an important step in computer aided diagnosis. In the present era, deep learning methods have achieved state of the art results for many medical image classification tasks such as diabetic retinopathy grading\cite{googledr,MahapatraGZSLTMI,LieTMI_2022,Devika_IEEE,MonusacTMI,Mahapatra_Thesis,KuanarVC,MahapatraTMI2021,JuJbhi2020,Frontiers2020,Mahapatra_PR2020}, digital patholology image classification \cite{googleDP,ZGe_MTA2019,Behzad_PR2020,Mahapatra_CVIU2019,Mahapatra_CMIG2019,Mahapatra_LME_PR2017,Zilly_CMIG_2016,Mahapatra_SSLAL_CD_CMPB,Mahapatra_SSLAL_Pro_JMI,Mahapatra_LME_CVIU,LiTMI_2015,MahapatraJDI_Cardiac_FSL,Mahapatra_JSTSP2014,MahapatraTIP_RF2014} and chest  xray images \cite{chexpert,NIHXray,MahapatraTBME_Pro2014,MahapatraTMI_CD2013,MahapatraJDICD2013,MahapatraJDIMutCont2013,MahapatraJDIGCSP2013,MahapatraJDIJSGR2013,MahapatraTrack_Book,MahapatraJDISkull2012,MahapatraTIP2012,MahapatraTBME2011,MahapatraEURASIP2010,MahapatraTh2012,MahapatraRegBook}, to name a few. Fully supervised learning (FSL)  methods that achieve state of the art results have access to disease classes (labels) in the training and test sets. 
However in many real-world scenarios we may not have access to samples of all possible diseases. A common scenario is the diagnosis of radiological images, such as chest xrays. 
Unseen classes are generally classified into one of the seen classes, resulting in wrong diagnosis and treatment planning. For deployment in clinical settings it is essential that a machine learning model learns to recognize novel test cases.

Zero shot learning (ZSL) aims to learn plausible representations of unseen classes from labeled data of seen classes, and recognize  unseen classes during test time. In a more generalized setting we expect to encounter both seen and unseen classes during the test phase, and a reliable model should accurately predict both classes. This is a case of generalized zero shot learning (GZSL) which is a challenging scenario since we do not want to predict unseen classes as one of the seen classes. 
We propose a GZSL method for medical image classification using self supervised learning (SSL), demonstrate its effectiveness across different datasets and also shows it's applicability to natural images. 


GZSL has been a widely explored topic for natural images \cite{FelixEccv18,VermaCVPR18,XianCVPR19,Mahapatra_CVAMD2021,PandeyiMIMIC2021,SrivastavaFAIR2021,Mahapatra_DART21b,Mahapatra_DART21a,LieMiccai21,TongDART20,Mahapatra_MICCAI20,Behzad_MICCAI20,Mahapatra_CVPR2020,Kuanar_ICIP19,Bozorgtabar_ICCV19,Xing_MICCAI19,Mahapatra_ISBI19,MahapatraAL_MICCAI18,Mahapatra_MLMI18} where seen and unseen classes are characterized by class attribute vectors. A model learns to correlate between class attribute vectors and  corresponding feature representations. This gives a strong reference point in synthesizing features of both seen and unseen classes, since by inputting the class attribute vector of the desired class the corresponding feature representation can be generated.
However medical images do not have such well defined class attributes since it requires high clinical expertise and time to define unambiguous attribute vectors for different disease classes. Hence it is not a straightforward task to apply state of the art GZSL methods to medical image classification. While this makes GZSL for medical images a challenging task, it is nevertheless essential to tackle this problem due to the potentially immense benefit.



Initial approaches to tackle ZSL \cite{Wu4,Manc47,Sedai_OMIA18,Sedai_MLMI18,MahapatraGAN_ISBI18,Sedai_MICCAI17,Mahapatra_MICCAI17,Roy_ISBI17,Roy_DICTA16,Tennakoon_OMIA16,Sedai_OMIA16,Mahapatra_OMIA16,Mahapatra_MLMI16,Sedai_EMBC16,Mahapatra_EMBC16,Mahapatra_MLMI15_Optic,Mahapatra_MLMI15_Prostate,Mahapatra_OMIA15,MahapatraISBI15_Optic,MahapatraISBI15_JSGR,MahapatraISBI15_CD,KuangAMM14} learnt cross-modal relationships between visual feature and semantic embeddings (class attribute vectors).
Subsequently, recent generative approaches to GZSL
\cite{Manc48,FelixEccv18,Mahapatra_ABD2014,Schuffler_ABD2014,Schuffler_ABD2014_2,MahapatraISBI_CD2014,MahapatraMICCAI_CD2013,Schuffler_ABD2013,MahapatraProISBI13,MahapatraRVISBI13,MahapatraWssISBI13,MahapatraCDFssISBI13,MahapatraCDSPIE13,MahapatraABD12,MahapatraMLMI12,MahapatraSTACOM12,VosEMBC,MahapatraGRSPIE12}, used generative adversarial networks (GANs) to optimize
the divergence between the data distribution of seen
classes and generated features. Consequently, generators trained
on seen class features cannot accurately represent unseen classes. The sub-optimal synthetic data does not lead to high performance of such models. 
As an attempt to circumvent this problem some methods \cite{Wu27,Wu22,MahapatraMiccaiIAHBD11,MahapatraMiccai11,MahapatraMiccai10,MahapatraICIP10,MahapatraICDIP10a,MahapatraICDIP10b,MahapatraMiccai08,MahapatraISBI08,MahapatraICME08,MahapatraICBME08_Retrieve,MahapatraICBME08_Sal,MahapatraSPIE08,MahapatraICIT06}
utilize unlabeled data of unseen classes in a transductive way. 
However they require two GANs for seen and unseen classes as they do not consider the relations between source and target domains. 

However, methods leveraging transductive approaches are particularly relevant for medical classification tasks \cite{WuCvpr20,ISR_MIDL_Ar,GCN_MIDL_Ar,DevikaAccess_Ar,SouryaISBI_Ar,Covi19_Ar,DARTGZSL_Ar,DARTSyn_Ar,Kuanar_AR2,TMI2021_Ar,Kuanar_AR1,Lie_AR2,Lie_AR,Salad_AR,Stain_AR,DART2020_Ar,CVPR2020_Ar,sZoom_Ar,CVIU_Ar,AMD_OCT,GANReg2_Ar,GANReg1_Ar,PGAN_Ar,Haze_Ar}. Absence of any supervised information from the unseen domain makes it very challenging to differentiate between disease labels, especially when many labels show similar appearance to the untrained eye. In our method we also leverage the unlabeled data of unseen classes as a guidance to train our GZSL method. 

Another tricky issue facing GZSL applications in general and medical images in particular is the potentially large semantic gap between images of different classes. Consequently synthesizing such unseen class features from the seen classes can be challenging. Leveraging unlabeled unseen class data (e.g., using anchors) can be effective in bridging the semantic gap \cite{WuCvpr20,Xr_Ar,RegGan_Ar,ISR_Ar,LME_Ar,Misc,Health_p,Pat2,Pat3,Pat4,Pat5}.
In an attemp to address the above challenges our paper makes the following contributions:
\begin{enumerate}
\item We propose a GZSL approach using self supervised learning (SSL) for medical image classification. Our method outperforms state of the art methods for multiple medical image datasets, and matches their performance on natural images.

    \item We use SSL for: 1) deriving anchor vectors through \textbf{improved} clustering; and 2) feature synthesis of seen and unseen classes.
    
    
    \item We achieve GZSL of medical images without using class attribute vectors commonly used for natural images. This is important for real world clinical scenarios where defining class attribute vectors is a time consuming and expensive task.  
    
     
\end{enumerate}

\section{Prior Work}


%


\textbf{(Generalized) Zero-Shot Learning:}
In Zero-Shot Learning \cite{Manc47,Pat6,Pat7,Pat8,Pat9,Pat10,Pat11,Pat12,Pat13,Pat14,Pat15,Pat16,Pat17,Pat18}, the goal is to recognize classes not encountered
during training. External information about the novel classes may be provided in
forms of semantic attributes \cite{Manc17}, visual descriptions \cite{Manc2}, or word embeddings \cite{Manc27}.
 Zero-shot learning has been addressed using Generative Adversarial
Networks (GANs) \cite{Manc48}, Variational Autoencoders (VAE) \cite{Manc38} or both of them \cite{XianCVPR19}.

In generalized zero-shot learning (GZSL), the purpose is to recognize images from known and unknown domains. Many works \cite{FelixEccv18,VermaCVPR18,Manc38,Manc48,XianCVPR19} obtain impressive results by training GANs in the known domain and generate unseen visual features from the semantic labels. This allows them to train a fully supervised classifier for two domains, which is robust to the biased recognition problem.
The work by Huang et al. \cite{HuangCVPR19} describes a Generative Dual
Adversarial Network (GDAN) which couples a Generator, a
Regressor and a Discriminator. The interaction between the three components produces various
visual features conditioned on class labels. 
Keshari et al. \cite{KeshariCvpr20} use overcomplete distributions to generate features of the unseen classes, while Min et al. \cite{MinCVPR20} use domain aware visual bias elimination for synthetic feature generation. Different from the above works we achieve GZSL without the need for descriptive class attribute vectors, but by specifying the class label of the desired output feature. GZSL for medical image tasks have seen limited applications such as registration \cite{Kori} and artefact reduction \cite{ChenISBI}.



\textbf{Self-Supervised Learning:}
These methods consist of two main approaches; 1) pre-text tasks and 2) down-stream tasks. Solving pre-text tasks learns a proper data representation, although the task itself may not be relevant, while down-stream tasks are used to evaluate the quality of features learned by self-supervised learning and are independent of pre-text tasks. Contrastive learning approaches such as MoCo \cite{HeMoco} and SimCLR \cite{ChenSimCLR} are popular and give state-of-the-art results for down-stream task-based methods. 
Self-supervised learning (SSL) also addresses labeled data shortage and has found wide use in medical image analysis by using innovative pre-text tasks for active learning \cite{MahapatraTMI2021}, anomaly detection \cite{Behzad_MICCAI20}, data augmentation \cite{Mahapatra_CVPR2020}, semi-supervised histology classification \cite{LuMahmood},stain normalization \cite{Mahapatra_MICCAI20} and registration \cite{TongDART20}.
%
Recent works also use self supervision for domain adaptation \cite{Saito} and perhaps the first work to combine GZSL and SSL \cite{WuCvpr20}. While our work is inspired from \cite{WuCvpr20} in using SSL for GZSL, and using GANs for feature synthesis, there are significant differences such as: 1) we do not use  class attribute vectors for training. Since medical images do not have defined class attribute vectors we use a simpler architecture for GZSL. 2) \cite{WuCvpr20} use a single generator but two discriminators to differentiate between seen and unseen classes. However we make use of a single generator and one discriminator to differentiate between all classes by leveraging anchor vectors; 3) We use a SSL based clustering approach to derive the anchor vectors of each class, including unseen classes. We use high level knowledge of the number of classes as a supervisory signal.

\section{Method}
\label{sec:method}

\begin{figure*}[t]
    \centering
    \includegraphics[height=5.7cm, width=16.8cm]{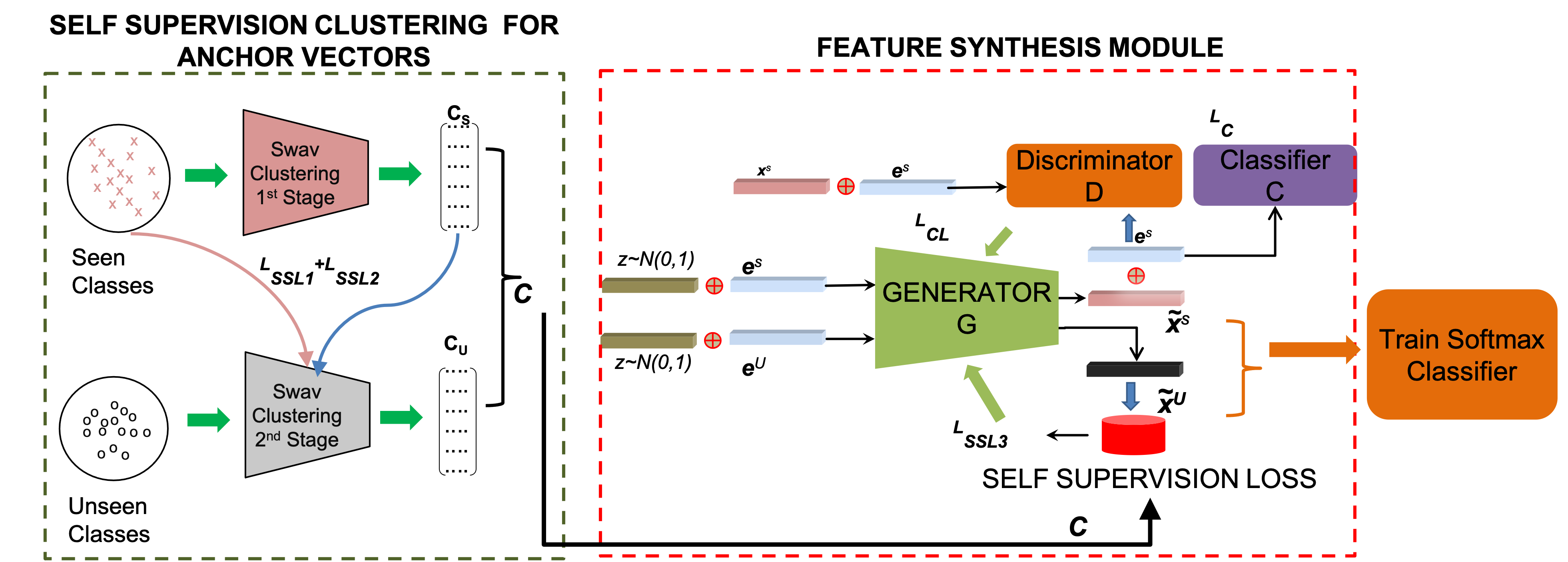}
    \caption{Architecture of proposed SC-GZSL method. In the first step we generate anchor vectors (cluster centroids) by using SSL within the SwAV clustering approach \cite{swav}. We have two clustering stages: one for Seen class samples and second for Unseen classes. Feature generation leverages one Generator and one Discriminator alongwith anchor vectors (from clustering) to derive SSL loss terms. }    \label{fig:workflow}
\end{figure*}%

\subsection{ Method Overview}

Figure~\ref{fig:workflow} depicts our proposed workflow. In the first step we generate anchor vectors (cluster centroids) by using SSL within the SwAV clustering approach \cite{swav}. We have two clustering stages: one for Seen class samples and second for Unseen classes. Anchor vectors of the Seen class samples are used to get SSL based loss terms for the second clustering stage. The second step involves feature generation that takes a noise vector and desired class label of output vector to synthesize features. Anchor vectors from the clustering stage are used to derive SSL based loss terms. Synthesized and real features of  unseen and seen classes are used to train a softmax classifier for identifying different disease classes.

\subsection{SSL  Clustering To Obtain Anchor Vectors}

Let the number of classes in the Seen set be $n_S$, and the number of classes in the Unseen set is $n_U$. We assume that the total number of classes is known.
We learn anchor vectors of the different classes by using the SSL based online clustering approach SwAV (\textbf{Sw}apping \textbf{A}ssignments between multiple \textbf{V}iews) \cite{swav}, and introduce additional SSL inspired loss terms. Typical offline clustering methods \cite{Swav2,Swav7} alternate between cluster assignment and centroid update. Since they require multiple passes over the dataset, such methods are slow for online clustering.
To overcome the high training time and inspired by contrastive instance learning \cite{Swav58}, \cite{swav} enforce that different augmentations of the same image are mapped to the same cluster. Multiple image views are contrasted by comparing their cluster assignments instead of features.

We take the cluster centers to be class anchor vectors since they give a reliable representation of the corresponding class. 
We choose to compute the anchor vectors in an online fashion since the number of unseen classes may change in a dynamic way depending upon the specific use case.
%
Given  image features $x_t$ and $x_s$ from two
different transformations of the same image, we compute their cluster assignments $q_t$ and $q_s$ by computing the distance of  the features to a set of $K$ cluster centers ${c_1,\cdots,c_K }$. A ``swapped'' prediction problem is solved with the following loss function:
\begin{equation}
    \mathcal{L}(x_t,x_s)=\ell (x_t,q_s)+\ell (x_s,q_t)
    \label{eq:l1}
\end{equation}
where $\ell(x,q)$ measures the fit between features $x$ and assignment $q$. Thus we compare features $x_t$ and $x_s$ using their intermediate cluster assignments $q_t$ and $q_s$ . If the two $x$'s capture same information, we can predict the cluster assignment from the other feature. 

\textbf{Online clustering:}
Given image $I_n$, it is transformed to $I_{nt}$ using transformation $t$  from a set $T$ of image transformations. A non-linear mapping $f_{\theta}$ transforms $I_{nt}$ to a feature vector which is projected to the unit sphere, i.e., $x_{nt} =
f_{\theta}(x_{nt})/\left\|f_{\theta}(x_{nt})\right\|_2$. The cluster assignment $q_{nt}$ is computed by determining the distance of  $x_{nt}$ to the set of cluster centroids, ${c_1,\cdots,c_K}$. $C$ denotes a matrix whose columns are $c_1,\cdots,c_k$. 

\paragraph{Swapped prediction problem:} 
%
Each term in Eq.\ref{eq:l1} represents the cross entropy loss between $q$ and the probability obtained by taking a softmax of the dot products of $x_i$ and all columns in $C$, i.e.,
\begin{equation}
    \ell (x_t,q_s)=-\sum_k q_s^{(k)} \log p_t^{(k)},~~ p_t^{(k)} =\frac{\exp{\frac{x_t^{\top}c_k}{\tau}}}{\sum_{k'} \exp{\frac{x_t^{\top}c_k}{\tau}} } 
    \label{eq:l2}
\end{equation}
where $\tau=0.1$ is the temperature parameter \cite{Swav58}. Computing this loss over all images and augmentations results in the following loss function for swapped prediction:
%
\begin{equation}
\begin{split}
 \mathcal{L}(x_t,x_s)= -\frac{1}{N} \sum_{n=1}^{N} \sum_{s,t\sim T} \biggl[ \frac{x_{nt}^{\top}Cq_{ns}}{\tau} + \frac{x_{ns}^{\top}Cq_{nt}}{\tau} \\ -\log\sum_{k=1}^{K} \exp\left(\frac{x_{nt}^{\top}c_k}{\tau} \right) -\log\sum_{k=1}^{K} \exp\left(\frac{x_{ns}^{\top}c_k}{\tau} \right) \biggr].
\end{split}
\label{eq:total}
\end{equation}

This loss function is jointly minimized with respect to the centroids in $C$ and  parameters $\theta$ of $f_{\theta}$.

\textbf{Computing the cluster assignments:} 
%
The clustering assignments $q$ are computed in an online fashion using image features within a batch. Since the centroids in $C$ are used across different batches, SwAV clusters multiple instances to their appropriate clusters. 
%
Given feature vectors $X=[x_1,\cdots,x_B ]$, we map them to  centroids  $C = [c_1 , \cdots, c_K ]$ using $Q = [q_1,\cdots, q_B]$, and we  optimize $Q$ to maximize the similarity between $X$ and $C$,
\begin{equation}
    \max_{Q\in\mathcal{Q}} Tr(Q^{\top}C^{\top}X) + \epsilon H(Q),
\end{equation}
where H is the entropy function, $H(Q) = -\sum_{ij} Q_{ij} \log Q_{ij}$ and $\epsilon=0.05$ controls  smoothness of mapping. A high $\epsilon$ could potentially  results in a trivial solution where all samples collapse into an unique representation and are assigned uniformly to all prototypes.  

%

\paragraph{Our Novel Contribution:}
We use the concept of anchor vectors to bridge the gap between seen and unseen classes, which is determined by the following steps:
Assuming we have $n_S$ seen classes we first cluster the Seen class images into $n_S$ clusters and obtain their centroids as $C_S={c_1,\cdots,c_{n_S}}$. In the next pass we compute the clusters $C_U={c_{n_S+1},\cdots,c_{n_S+n_U}}$ of the $n_U$ unseen classes using the following additional constraints:
\begin{enumerate}
    \item The centroids in $C_S$ do not change since they have been computed from the seen classes.
    \item A  self supervised constraint is added where the centroids of the unseen classes are forced to be different from the seen class centroids. This is done to account for the situation that some of the Unseen classes may be semantically close to one or more Seen classes. This may happen when images of different disease labels have very similar appearance which can be a common occurrence for radiological images. This condition is implemented using:
    \begin{equation}
        \mathcal{L}_{SSL1}=\min\left(CoSim (C_S^i,C_U^j),\sigma_1\right)
        \label{eq:lssl1}
    \end{equation}
    Here $\sigma_1=0.15$ is a parameter that determines the semantic distance between the centroids, and CoSim denotes cosine similarity. 
    
    \item We add a second self supervised constraint that the similarity of seen class sample, $x_s^i$, with its corresponding class centroid $C_S^i$ is higher than their similarity w.r.t all $C_U^j$. This is achieved by randomly selecting samples from the Seen class training set during minibatch training and computing the different cosine similarities. This constraint is implemented by 
    \begin{equation}
    \begin{split}
        & \mathcal{L}_{SSL2}= \\ 
        & \max\left(CoSim(x_S^i,C_S^i)-CoSim(x_S^i,C_U^j),\sigma_2\right) \forall j 
        \label{eq:lssl2}
        \end{split}
    \end{equation}
    $\sigma_2=0.25$ controls the minimum degree of semantic difference between different classes.
 
\end{enumerate}

The final loss term for clustering the Unseen class samples is $\mathcal{L}_{Unseen}=\mathcal{L}(x_s,x_t)+\lambda_{1}L_{SSL1}-\lambda_{2}L_{SSL2}$, where $\mathcal{L}(x_s,x_t)$ is defined in Eqn.~\ref{eq:total}. $\lambda_{1}=1.1,\lambda_{2}=0.7$ are the weights. The `$-\lambda_{2}L_{SSL2}$' ensures that the loss term does not increase arbitrarily which is possible for `$+\lambda_{2}L_{SSL2}$'.

\subsection{Feature Generation Network}

Given the training images of Seen classes and unlabeled images of the Unseen classes we learn a generator $G : \mathcal{E},\mathcal{Z} \longrightarrow \mathcal{X}$ , which takes a class label vector $e^{y} \in \mathcal{E}$
and a Gaussian noise vector $z \in \mathcal{Z}$ as inputs, and generates a 
 feature vector $\tilde{x} \in \mathcal{X}$. The discriminator $D:\mathcal{X},\mathcal{E} \xrightarrow{}[0,1]$ takes a real feature $x$
or synthetic feature $\tilde{x}$ and corresponding class label vector $e^{y}$ as input and determines whether the feature vector matches the class label vector. 
The generator $G$ aims to fool $D$ by producing features highly correlated with
$e^{y}$ using a Wasserstein adversarial loss\cite{Wu3}:
\begin{equation}
    \begin{split}
        \mathcal{L}_{WGAN}=\min_G \max_D \mathbb{E}[D(x,e^y)]-\mathbb{E}[D(\tilde{x},e^y)] \\
        -\lambda \mathbb{E}[(\left\|\nabla_{\tilde{x}} D(\tilde{x},e^{y}) \right\|_2-1)^2]
    \end{split}
    \label{eq:wgan}
\end{equation}
where the third term is a gradient penalty term, and $\tilde{x}=\alpha x + (1-\alpha)\tilde{x}$. $\alpha\sim U(0,1)$ is sampled from a uniform distribution.


\subsubsection{Self Supervised Loss From Anchor Vectors}
\label{met:synth}

The discriminator $D$ is a classifier that determines whether the generated feature vector $\tilde{x}$ belongs to one of the seen classes. Since the unseen classes are not labeled we do not have a data distribution for them and hence we use self supervision to determine whether the generated feature vector matches an unseen class. 
As the anchor vectors (i.e., the cluster centers) are fixed, we calculate the cosine distance between the generated vector $\tilde{x}$ and the anchor vector corresponding to the desired class $y$, i.e.
\begin{equation}
    \mathcal{L}_{SSL3} =1-CoSim(\tilde{x},c_y)
    \label{eq:lssl3}
\end{equation}
If $\tilde{x}$ truly represents the desired class $y$ then the cosine similarity between $\tilde{x}$ and the corresponding anchor vector $c_y$ should be highest amongst all $K(=n_S+n_U)$ anchor vectors, and the corresponding loss is lowest. 
%



\subsubsection{Classifier Loss}

We expect that $\tilde{x}^{s}$ (synthesized feature vector for seen classes) are predicted correctly by  a pre-trained classifier $CL$ with a loss defined as below
\begin{equation}
    \mathcal{L}_{CL}=-\mathbb{E}_{(\tilde{x}^{s},y^{s})\sim P_{\tilde{x}^{s}}} \left[\log P(y^{s}|\tilde{x}^{s},\theta_{CL}) \right]
\end{equation}
where $P(y^{s}|\tilde{x}^{s},\theta_{CL})$ is the classification probability and $\theta_{CL}$ denotes fixed parameters of the pre-trained classifier.

\subsection{Training and Implementation}

The final loss function is defined as 
\begin{equation}
\mathcal{L}=\mathcal{L}_{WGAN} + \lambda_{CL}\mathcal{L}_{CL} + \lambda_{3}L_{SSL3} 
\end{equation}
where $\lambda_{CL},\lambda_{3}$ are weights that balance the contribution of the different terms. Once training is complete we specify the label of desired class and input a noise vector to $G$ which synthesizes a new feature vector. We combine the synthesized target features of the unseen class $\tilde{x}^{u}$ and real and synthetic features of seen class $x^{s},\tilde{x}^{s}$ to construct the training set. Then we train a softmax classifier by minimizing the negative log likelihood loss: 
\begin{equation}
    \min_{\theta} - \frac{1}{|\mathcal{X}|} \sum_{(x,y)\in(\mathcal{X},\mathcal{Y})} \log P(y|x,\theta),
\end{equation}
where $P(y|x,\theta)=\frac{\exp(\theta^{T}_y x)}{\sum_{j=1}^{|\mathcal{Y}|} \exp(\theta^{T}_y x)}$ is the classification probability
and $\theta$ denotes classifier parameters. The final class prediction is by $f(x)=\arg \max_y P(y|x,\theta)$
%

\textbf{Implementation Details:}
We show results for natural and medical images and compare with  existing GZSL methods. 
Extending our method to natural images is straightforward where in we replace the class label vector $e^y$ with the corresponding class attribute vectors.
For feature extraction, similar to \cite{Manc47}, we use a pre-trained ResNet-101 to extract $2048$ dimensional CNN features for natural images. 
The generator (G) and discriminator (D) are all multilayer perceptrons. $G$
has two hidden layers of $2000$ and $1000$ units respectively while the discriminator D is implemented with one hidden layer of $1000$ hidden
units. We choose Adam \cite{Adam} as our optimizer, and the momentum is set to ($0.9, 0.999$). The values of loss term weights are $\lambda_{CL}=0.6, \lambda_{3}=0.9$.
 Training the Swav Clustering algorithm takes $12$ hours and the feature synthesis network for $50$ epochs takes $17$ hours, all on a single NVIDIA V100 GPU (32 GB RAM). PyTorch was used for all implementations.


\subsection{ Evaluation Protocol}

%
The seen class $S$ can have samples from $2$  or more disease classes, and the unseen class $U$ contains samples from the remaining classes. We use all possible combinations of labels in $S$ and $U$.
Following standard practice for GZSL, average class accuracies are calculated for two settings:  1) $\textbf{S}$: training is performed on synthesized samples of $S+U$ classes and test on $S_{Te}$. 2) $\textbf{U}$: training is performed on synthesized samples of $S+U$ classes and test on $U$. We also report the harmonic mean defined as 
\begin{equation}
    H=\frac{2\times Acc_U \times Acc_S}{Acc_U + Acc_S}
\end{equation}
where $Acc_S$ and $Acc_U$ denote the accuracy of images
from seen (setting $S$) and unseen (setting $U$) classes respectively:


\section{Experimental Results}
\label{sec:expt}



\begin{figure*}[t]
 \centering
\begin{tabular}{cccc}
\includegraphics[height=3.3cm, width=4cm]{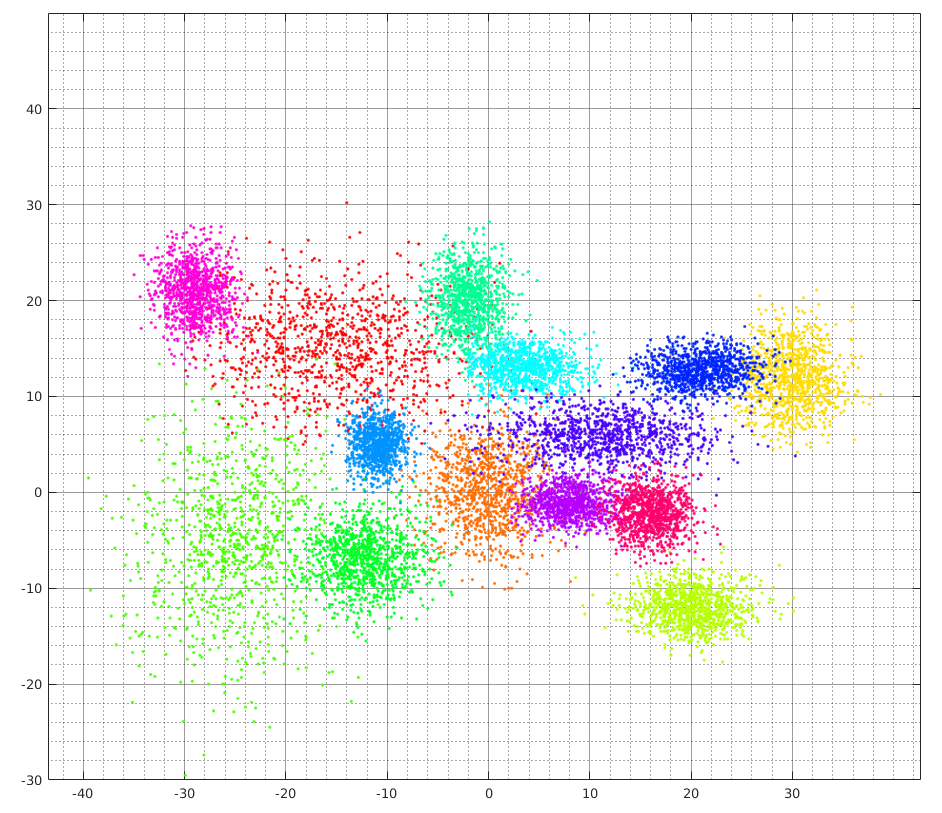} &
\includegraphics[height=3.3cm, width=4cm]{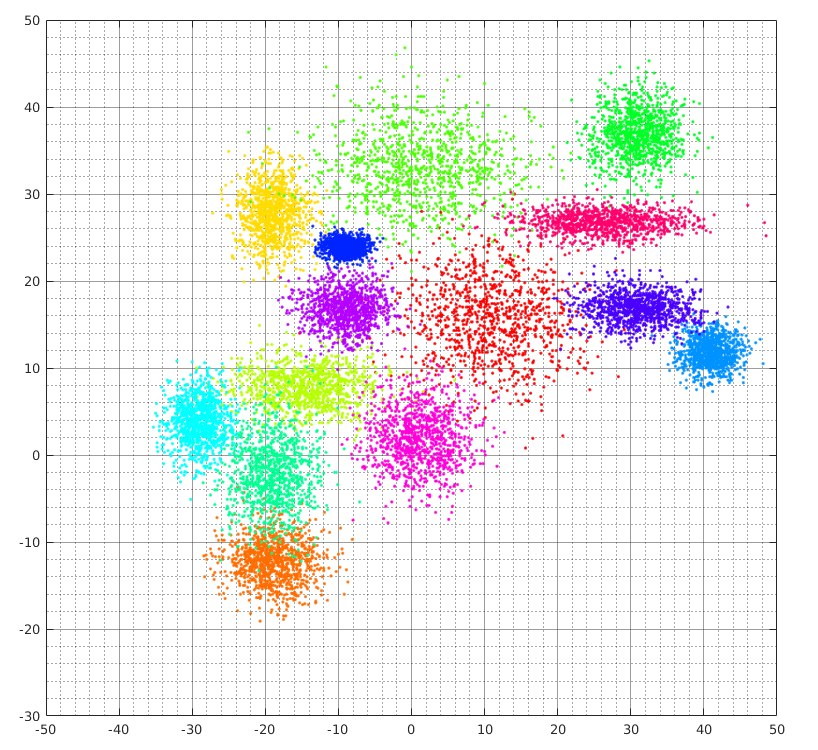} &
\includegraphics[height=3.3cm, width=4cm]{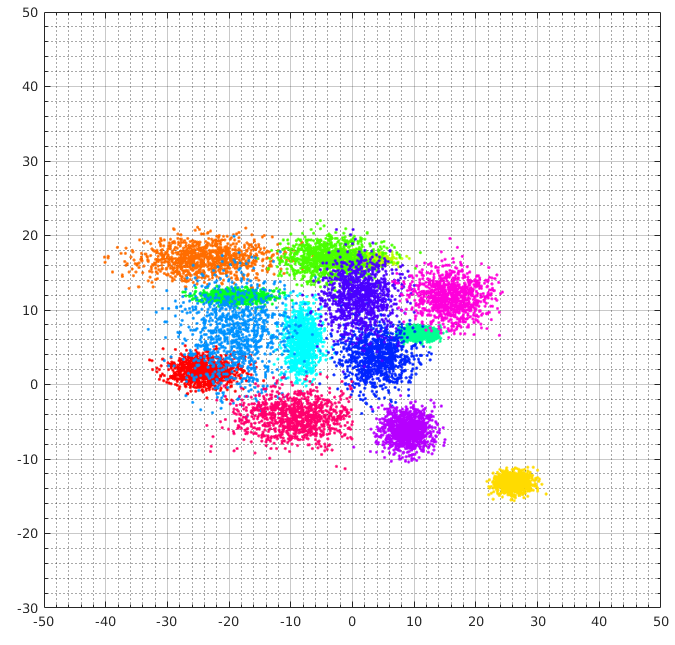} &
\includegraphics[height=3.3cm, width=4cm]{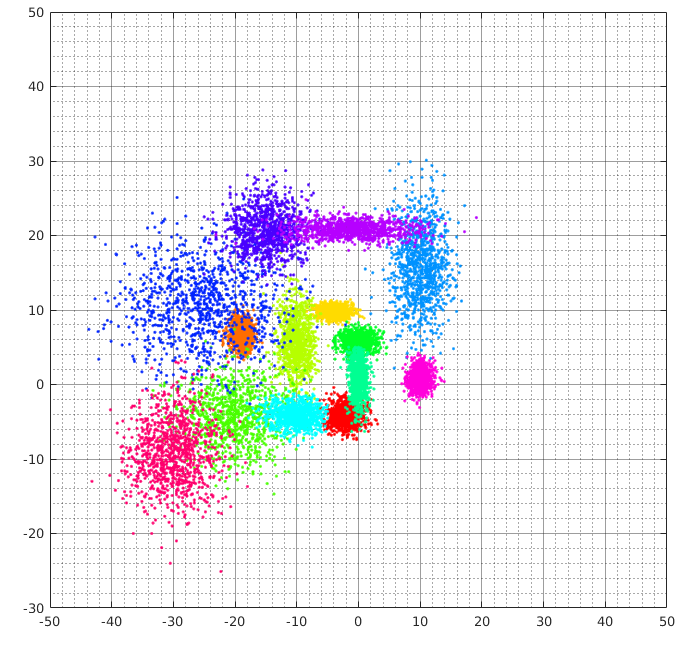} \\
(a) & (b) & (c) & (d) \\
\end{tabular}
\begin{tabular}{c}
\includegraphics[height=1.0cm, width=15cm]{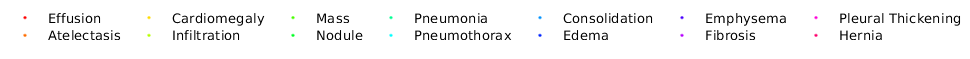}  \\
\end{tabular}
\caption{Feature visualizations for NIH ChestXray Dataset: (a) Seen+Unseen classes from actual dataset; distribution of synthetic samples generated by (b) SC-GZSL; (c) SC-GZSL$_{w\mathcal{L}_{SSL3}}$; (d) SDGN \cite{WuCvpr20}. Different colours represent different classes. (b) is closer to (a), while (c) and (d) are quite different.}
\label{fig:FeatVis}
\end{figure*}

\subsection{Dataset Description}

We demonstrate our method's effectiveness on natural images and the following medical imaging datasets for classification tasks. 
%
 Datasets with a minimum of 3 disease classes (excluding normal label) were chosen to highlight the performance of feature synthesis. 
\begin{enumerate}
    \item \textbf{CAMELYON17} dataset \cite{camelyon17}: contains 1000 whole slide images (WSIs) with 5 slides per patient: 500 slides for training and 500 slides for test. Training set has annotations of 3 categories of lymph node metastasis: Macro (Metastases greater than $2.0$ mm), Micro (metastasis greater than $0.2$ mm or more than $200$ cells, but smaller than $2.0$ mm), and ITC (single tumor cells or a cluster of tumor cells smaller than $0.2$mm or less than $200$ cells). We extract $224\times224$ patches from the different slides and obtain $130,000$ tumor patches and $200,000$ normal patches. We take a pre-trained ResNet101 and finetune the last FC layer using the CAMELYON16 dataset \cite{camelyon16}, which is closely related but different from CAMELYON17. A baseline fully supervised learning (FSL) method is implemented\footnote{https://grand-challenge-public.s3.amazonaws.com/evaluation-supplementary/80/46fc579c-51f0-40c4-bd1a-7c28e8033f33/Camelyon17_.pdf} which is the top ranked in the leaderboard.

\item \textbf{NIH Chest Xray} Dataset: For lung disease classification we adopted the NIH ChestXray14 dataset \cite{NIHXray} having $112,120$ expert-annotated frontal-view X-rays from $30,805$ unique patients and has $14$ disease labels. Original images were resized to $224\times224$. A pre-trained resnet-101 was finetuned using the CheXpert dataset \cite{chexpert} and the chosen baseline FSL was from \cite{CheXNet}.

\item \textbf{CheXpert} Dataset: We used the CheXpert dataset \cite{chexpert} consisting of $224,316$ chest radiographs of $65,240$ patients labeled for the presence of $14$ common chest conditions. 
Original images were resized to $224\times224$.
A pre-trained resnet-101 was finetuned using the NIH dataset \cite{NIHXray} and the baseline FSL method was of \cite{CXtop2} which is ranked second for the dataset with shared code.

\item \textbf{Kaggle Diabetic Retinopathy} dataset: has approximately $35,000$ images in the provided training set \cite{KaggleDR}. 
Images are labeled by a single clinician with the respective DR grade, out of 4 severity levels: 1- mild(2443 images), 2-moderate (5291 images), 3-severe (873 images), and 4-proliferative DR (708 images). The normal class $0$ has 25810 images. A pre-trained resnet-101 was finetuned using \cite{DMR} which has $9939$ color fundus images ($2720\times2720$) from $2740$ diabetic patients. Although the number of classes are different from Kaggle the features are accurate since the end task is DR detection. The chosen baseline method was of \cite{AraujoDR}. Original images were resized to $224\times224$.


\item \textbf{Gleason grading challenge} dataset \footnote{https://gleason2019.grand-challenge.org/Home} for prostate cancer (PCA) \cite{GleasonData}.
 It has $333$  Tissue Microarrays (TMAs) from $231$ patients and has $5$ Gleason grades. Six pathologists with  $27, 15, 1, 24, 17$, and $5$ years of experience annotated the data and majority voting was used to construct the “ground truth
label”. The training set had $200$ TMAs while the validation set had $44$ TMAs. A separate test set consisting of $87$  TMAs from 60
other patients. Although a much larger dataset for PCA using WSIs is available\footnote{https://www.kaggle.com/c/prostate-cancer-grade-assessment/overview}, the data cannot be used for external submissions\footnote{https://www.kaggle.com/c/prostate-cancer-grade-assessment/discussion/201117}. The baseline FSL was the classification outcome of the top ranked method\footnote{https://github.com/hubutui/Gleason}. The feature extractor was a pre-trained ResNet101 finetuned using the CAMELYON16 dataset \cite{camelyon16}. Since both are histopathology image datasets, the feature extractor is quite accurate. The high dimensional images were divided into $224\times224$ patches. The individual labels patches from normal images were all `normal'. For the diseased images (all Gleason grades except 1), the labels of individual patches were obtained using the multiple instance learning method of \cite{GGL26}. Thus we obtained more than $5,000$ patches of each label.

\end{enumerate}
Since we did not have labels of the organizer designated test sets of all datasets, a $70/10/20$ split at patient level was done to get training, validation and test sets for NIH Chest Xray, CheXpert and Kaggle DR datasets.

For natural images we use the following  five datasets: 1) CUB \cite{Wu39}, AwA1 \cite{Manc17}, AwA2 \cite{Manc47},
SUN \cite{Wu26}, and FLO \cite{Wu24}. CUB includes 11K images and
200 species of birds labeled with 312-D attributes. AwA1
and AwA2 consist of 50 kinds of animals described by 85-
D attributes, containing 30K and 37K images respectively.
SUN is a large-scale scene attribute dataset, including 717
classes and 14K images with 102-D attributes. FLO con-
sists of 8K images from 102 flower classes. 
%
Adapting our method to natural images is done by replacing the class vector (e) with the class attribute vector.


\subsection{Baseline Methods}

We compare our method's performance with the following GZSL methods employing different feature generation approaches such as CVAE or GANs:
1) CVAE based generation method of \cite{HuangCVPR19}; 2) over complete  distribution (OCD) method of \cite{KeshariCvpr20}; 3) self-supervised learning GZSL method of \cite{WuCvpr20}; 
4) FSL- Top performing FSL methods of corresponding datasets.
 Following GZSL protocol we report performance for Seen and Unseen classes. 
 Our method is denoted as SC-GZSL (\textbf{S}elfsupervised \textbf{C}lustering based \textbf{GZSL}).


\subsection{Visualization of Synthetic Image Features}

Figure~\ref{fig:FeatVis} (a) shows t-SNE plot of features from actual data from the NIH chest Xray dataset where the different classes are spread over a wide area, with slight overlap between some classes. Figure~\ref{fig:FeatVis} (b) shows the distribution of synthetic features generated by our method. Although the corresponding clusters for the different classes have separate locations in the two figures they are similar to that of Figure~\ref{fig:FeatVis} (a) in the sense that the different classes are similarly separated.
Figure~\ref{fig:FeatVis} (c) shows the feature distribution for our method without using self-supervision. The resulting distribution is compact without overlap between classes, which is not representative of the real-world case. Classifiers trained on such distributions perform poorly on unseen classes. Figure~\ref{fig:FeatVis} (d) shows the feature distributions using SDGN \cite{WuCvpr20}. Although it also uses SSL the resulting feature representation is less accurate than our proposed method which contributes to the corresponding inferior performance. 

\subsection{Generalized Zero Shot Learning Results}

Table \ref{tab:GZSL_Natural} summarizes the results of our algorithm on natural images. The best performing method amongst all competing methods is SDGN \cite{WuCvpr20}. However we are able to outperform it despite using a much simpler architecture. A McNemar's statistical test \cite{Keshari21} shows that the results between SC-GZSL and SDGN is not very significant ($p=0.062$), except for the SUN dataset (p=0.01). This dataset is particularly challenging as demonstrated by the fact that accuracy values are lower than other datasets.

The results for medical images shown in Table~\ref{tab:GZSL_Medical} shows our proposed method outperforms all competing GZSL methods including SDGN. This significant difference in performance can be explained by the fact that the complex architectures that worked for natural images will not be equally effective for medical images which have less information. Absence of attribute vectors for medical images is another contributing factor. The class attributes provide a rich source of information about natural images which can be leveraged using existing architectures. On the other hand medical images require a different approach.

\begin{table*}[t]
 \begin{center}
\begin{adjustbox}{width=\textwidth}
\begin{tabular}{|c|ccc|ccc|ccc|ccc|ccc|}
\hline 
{} & \multicolumn{15}{|c|}{\textbf{Natural Images}} \\ \cline{2-16}
{Method} & \multicolumn{3}{|c|}{CUB} & \multicolumn{3}{|c|}{AwA1} & \multicolumn{3}{|c|}{AwA2} & \multicolumn{3}{|c|}{SUN} & \multicolumn{3}{|c|}{FLO} \\ \cline{2-16}
{} & S & U & H & S & U & H & S & U & H & S & U & H & S & U & H \\ \hline
f-Vaegan \cite{XianCVPR19} & 65.1 & 61.4 & 63.2 & - & - & - & 88.6 & 84.8 & 86.7 & 41.9 & 60.6 & 49.6 & 87.2 & 78.7 & 82.7 \\ \hline
GXE \cite{Wu18} & 68.7 & 57.0 & 62.3 & 89.0 & 87.7 & 88.4 & 90.0 & 80.2 & 84.8 & 58.1 & 45.4 & 51.0 & - & - & - \\ \hline
SDGN \cite{WuCvpr20} & 70.2 & 69.9 & 70.1 & 88.1 & 87.3 & 87.7 & 89.3 & 88.8 & 89.1 & 46.0 & 62.0 & 52.8 & 91.4 & 78.3 & 84.4 \\ \hline
GDAN \cite{HuangCVPR19} & 66.7 & 39.3 & 49.5 & - & - & - & 67.5 & 32.1 & 43.5 & 89.9 & 38.1 & 53.4 & - & - & - \\ \hline
OCD\cite{KeshariCvpr20} & 59.9 & 44.8 & 51.3 & - & - & - & 73.4 & 59.5 & 65.7 & 42.9 & 44.8 & 43.8 & - & - & - \\ \hline
{SCGZSL} & 71.7 & 70.6 & 71.1 & 88.5 & 88.1 & 88.3 & 89.9 & 89.3 & 89.6 & 50.3 & 62.1 & 55.6 & 91.8 & 79.4 & 85.2 \\ \hline
\end{tabular}
\end{adjustbox}
\caption{\textbf{GZSL Results For Natural Images:}Average per-class classification accuracy ($\%$) and harmonic mean (H) accuracy of generalized zero-shot learning when test samples are from  Seen(S)  or Unseen (U) classes. Numbers for competing methods are taken from \cite{WuCvpr20}. $S,U$ denote $Acc_S,Acc_U$. }
\label{tab:GZSL_Natural}
\end{center}
\end{table*}

\begin{table*}[t]
 \begin{center}
\begin{adjustbox}{width=\textwidth}
\begin{tabular}{|c|ccc|ccc|ccc|ccc|ccc|}
\hline 
{} & \multicolumn{15}{|c|}{\textbf{Multiple Medical Image Datasets}} \\ \cline{2-16}
{Method} & \multicolumn{3}{|c|}{CAMELYON17} & \multicolumn{3}{|c|}{NIH Xray} & \multicolumn{3}{|c|}{CheXpert} & \multicolumn{3}{|c|}{Kaggle DR} & \multicolumn{3}{|c|}{Gleason} \\ \cline{2-16}
{} & S & U & H & S & U & H & S & U & H & S & U & H & S & U & H \\ \hline
f-VAEGAN \cite{XianCVPR19} & 90.2 & 88.2 & 89.2 & 82.9 & 80.0 & 81.4 & 88.5 & 87.6 & 88.0 & 92.8 & 90.2 & 91.5 & 88.2 & 85.1 & 86.6  \\ \hline
GDAN \cite{HuangCVPR19} & 91.1 & 89.1 & 90.1 & 83.8 & 80.9 & 82.3 & 89.2 & 88.0 & 88.6 & 94.2 & 91.0 & 92.6 & 88.8 & 86 & 87.4 \\ \hline
OCD\cite{KeshariCvpr20} & 91.5 & 89.3 & 90.4 & 84.7 & 81.3 & 83.0 & 89.9 & 88.1 & 89.0 & 94.8 & 91.3 & 93.0 & 89.2 & 86.9 & 88 \\ \hline
SDGN \cite{WuCvpr20} & 92.1 & 89.5 & 90.8 & 84.4 & 81.1 & 82.7 & 90.2 & 88.2 & 89.2 & 95.0 & 91.9 & 93.4 & 90.0 & 87.8 & 88.9 \\ \hline
{SCGZSL} & 93.5 & 91.1 & 92.3 & 87.2 & 84.3 & 85.7 & 91.8 & 89.4 & 90.6 & 96.1 & 93.2 & 94.7 & 92.1 & 89.5 & 90.8 \\ \hline
{FSL} & 93.7 & 93.5 & 93.6 & 87.4 & 86.9 & 87.1 & 92.1 & 92.5 & 92.3 & 96.4 & 96.1 & 96.2 & 92.4 & 92.2 & 92.3 \\ \hline \hline
{SCGZSL} & \multicolumn{15}{|c|}{\textbf{Ablation Studies}} \\ \cline{2-16}
{$_{w\mathcal{L}_{SSL1}}$} & 91.2 & 88.7 & 89.9 & 84.5 & 82.1 & 83.3 & 89.1 & 86.9 & 88.0 & 92.2 & 89.6 & 90.9 & 90.3 & 86.9 & 88.6  \\ \hline
{$_{w\mathcal{L}_{SSL2}}$} & 90.8 & 88.1 & 89.4 & 84.0 & 82.2 & 83.1 & 88.8 & 86.2 & 87.5 & 91.8 & 88.2 & 90.0 & 89.2 & 86 & 87.6 \\ \hline
{$_{w\mathcal{L}_{SSL3}}$} & 90.0 & 87.0 & 88.5 & 83.2 & 81.0 & 82.1 & 87.6 & 85.1 & 86.3 & 90.1 & 86.7 & 88.4 & 88.4 & 85.5 & 86.9 \\ \hline
{only ${\mathcal{L}_{SSL3}}$} & 89.3 & 86.4 & 87.8 & 82.6 & 80.7 & 81.6 & 87.0 & 84.5 & 85.7 & 88.9 & 85.9 & 87.4 & 87.7 & 84.9 & 86.3 \\ \hline
{$_{\mathcal{L}}$} & 87.2 & 84.1 & 85.6 & 80.7 & 79.1 & 79.7 & 84.6 & 82.7 & 83.6 & 86.5 & 83.7 & 85.1 & 86.1 & 82.8 & 84.4 \\ \hline
\end{tabular}
\end{adjustbox}
\caption{\textbf{GZSL Results For Medical Images:}Average per-class classification accuracy ($\%$) and harmonic mean accuracy of generalized zero-shot learning when test samples are from  Seen (Setting $S$) or unseen (Setting $U$) classes. Results of ablation studies are also shown.}
\label{tab:GZSL_Medical}
\end{center}
\end{table*}

\subsection{Ablation Studies}

%
Table~\ref{tab:GZSL_Medical} also shows results for the following ablation studies: 1) SCGZSL$_{w\mathcal{L}_{SSL1}}$- SCGZSL without the loss term $\mathcal{L}_{SSL1}$ (Eqn.\ref{eq:lssl1}) for obtaining the anchor vectors, 2) SCGZSL$_{w\mathcal{L}_{SSL2}}$ - SCGZSL without the loss term $\mathcal{L}_{SSL2}$ (Eqn.\ref{eq:lssl2}) to get anchor vectors, 3) SCGZSL$_{\mathcal{L}}$- Using only the baseline loss term  $\mathcal{L}(z_s,z_t)$ (Eqn.\ref{eq:total}) for clustering all seen and unseen classes together, and no $\mathcal{L}_{SSL3}$ for feature synthesis; 4) SCGZSL$_{w\mathcal{L}_{SSL3}}$- SCGZSL without the loss term $\mathcal{L}_{SSL3}$ (Eqn.\ref{eq:lssl3}) for training the feature synthesis network; 5) SCGZSL$-\text{only} \mathcal{L}_{SSL3}$- SCGZSL using only $\mathcal{L}_{SSL3}$ for feature synthesis without $\mathcal{L}_{SSL1}$, $\mathcal{L}_{SSL2}$.

The first three ablation studies investigate the effect of clustering on the final classification results. Their significant performance degradation compared to SCGZSL indicates the importance of our novel SSL based terms ($\mathcal{L}_{SSL1},\mathcal{L}_{SSL2}$) in obtaining accurate anchor vectors. The baseline method, SCGZSL$_{\mathcal{L}}$, does not use any form of self supervision and has lowest $H$ values. Compared to SCGZSL, we observe that excluding  ${\mathcal{L}_{SSL3}}$ (SCGZSL$_{w\mathcal{L}_{SSL3}}$) leads to maximum reduction of $H$ (more than $3.5\%$) across all datasets . This indicates that ${\mathcal{L}_{SSL3}}$ makes the most significant contribution to our method's performance. The use of anchor vectors makes it easier to synthesize features of unseen classes.

The influence of ${\mathcal{L}_{SSL1}},{\mathcal{L}_{SSL2}}$ is quantitatively similar as shown by similar $H$ values of SCGZSL$_{w\mathcal{L}_{SSL1}}$, SCGZSL$_{w\mathcal{L}_{SSL2}}$ across all datasets. However their difference in $H$ values compared to SCGZSL is nearly $2.4\%$ which is significant ($p=0.01$). Thus the use of self supervision is an important factor in obtaining accurate anchor vectors (cluster centroids). Although the baseline clustering mechanism, SwAV, uses self supervision in the form of contrastive loss, including $\mathcal{L}_{SSL1}$ and $\mathcal{L}_{SSL2}$ sigificantly improves clustering accuracy. Excluding both $\mathcal{L}_{SSL1},\mathcal{L}_{SSL2}$ and using the baseline SwAV (`only ${w\mathcal{L}_{SSL3}}$')  gives significantly reduced $H$ values  for the different datasets   despite using $\mathcal{L}_{SSL3}$ for feature synthesis. This clearly indicates the importance of having accurate anchor vectors for our method. SCGZSL$_{\mathcal{L}}$ can be considered as the most basic method without using any of our proposed novel loss terms, and unsurprisingly gives the worst results.

\begin{figure}[h]
 \centering
\begin{tabular}{c}
\includegraphics[height=3.0cm, width=7.4cm]{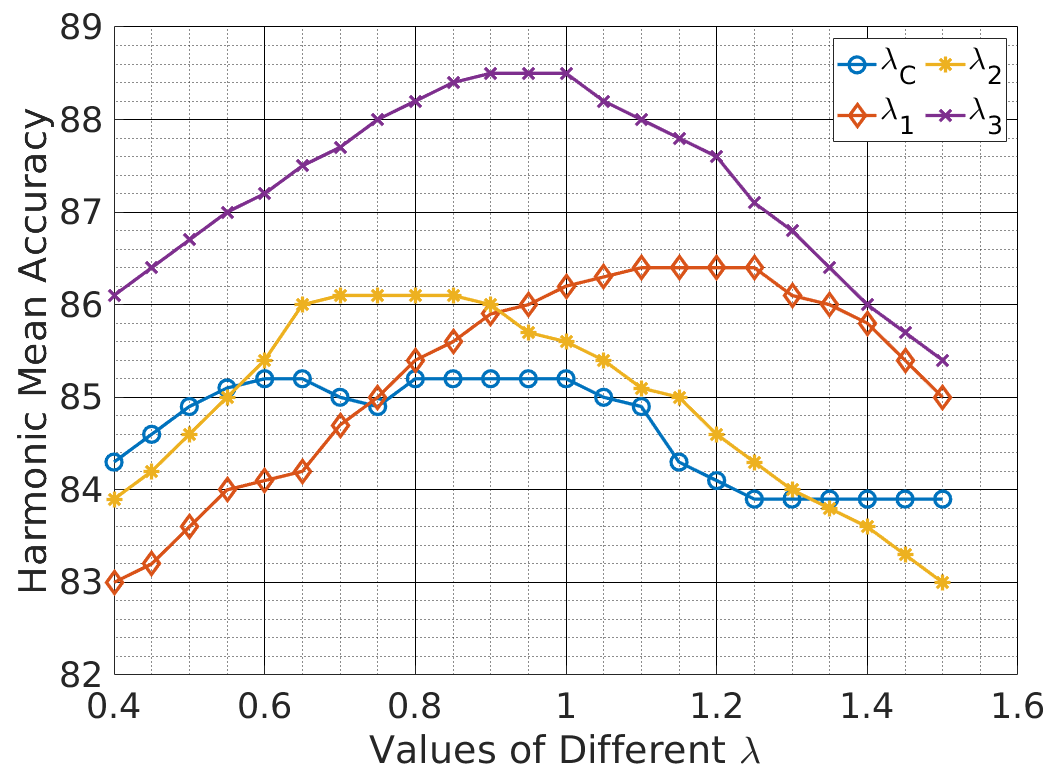} \\
(a) \\
\includegraphics[height=3.0cm, width=7.4cm]{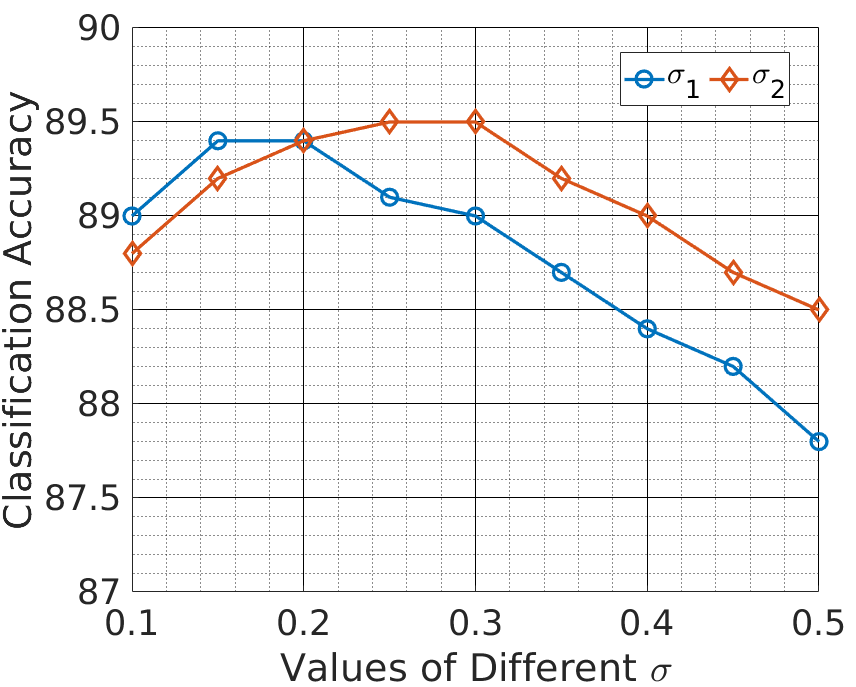}  \\
  (b) \\
\end{tabular}
\caption{Hyperparameter Plots showing the value of $H$ and classification accuracy for different values of;(a) $\lambda$; (b) $\sigma$.}
\label{fig:hyper}
\end{figure}

\subsection{Hyperparameter Selection}

For all the competing methods in the case of medical images we start with the original values provided by the authors and vary them in range $x\pm0.5x$ in steps of $x/10$, where $x$ is the initial value. The best results are usually obtained using author provided values for each method.

Figure~\ref{fig:hyper} (a) shows the harmonic mean values for the NIH Chest Xray dataset for different values of hyperparameters $\lambda_{1},\lambda_{2},\lambda_{3}$, while Figure~\ref{fig:hyper} (b) shows the corresponding plots for different values of $\sigma_1,\sigma_2$. The $\lambda$'s were varied between $[0.4-1.5]$ in steps of $0.05$ and the performance on a separate test set of $10,000$ images was monitored. We start with the base cost function of Eqn.~\ref{eq:wgan}, and first select the optimum value of $\lambda_1$. $\lambda_1$ values is fixed and we then determine $\lambda_2$, and then $\lambda_3$ by fixing $\lambda_1,\lambda_2$. The order in which the parameters were set is important and we find the above order as giving the best results.
Similarly the value of $\sigma$'s were varied between $[0.1,0.5]$ in steps of $0.05$, and the resulting classification accuracy of the Xray images was determined. i.e., whether they were assigned to the correct cluster (class). 
%



\begin{figure}[t]
 \centering
\begin{tabular}{cc}
\includegraphics[height=1.8cm, width=3.7cm]{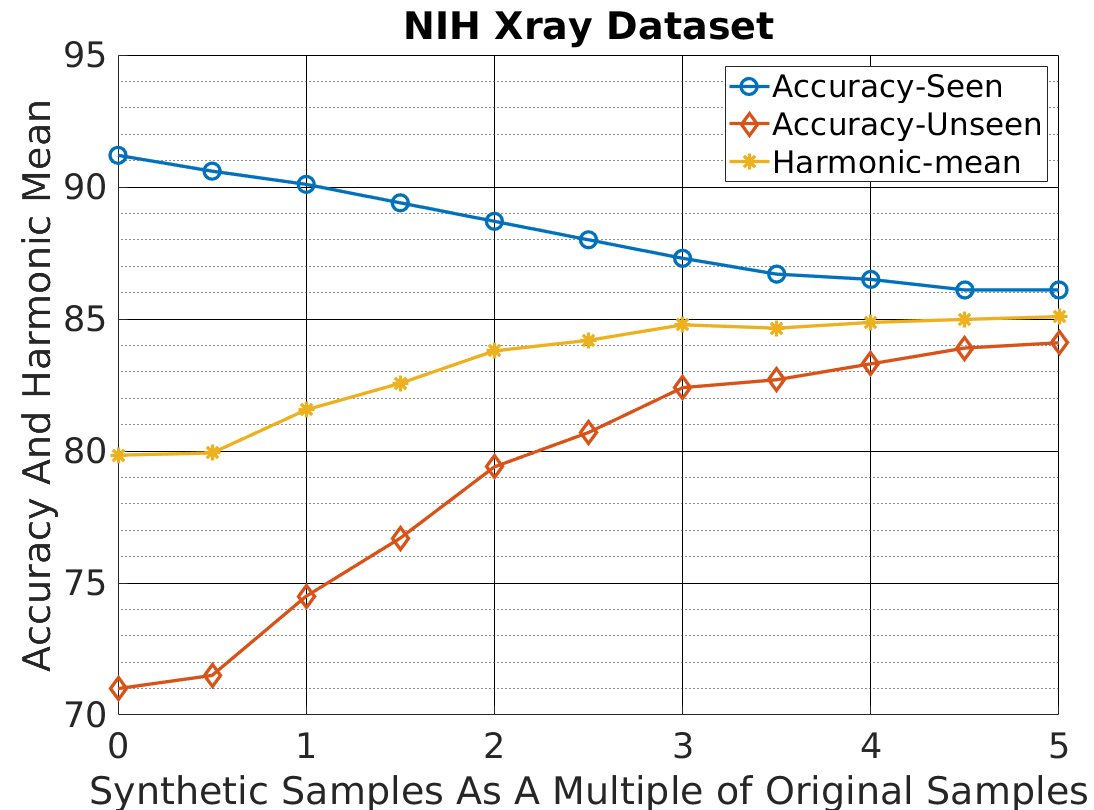} &
\includegraphics[height=2.0cm, width=3.7cm]{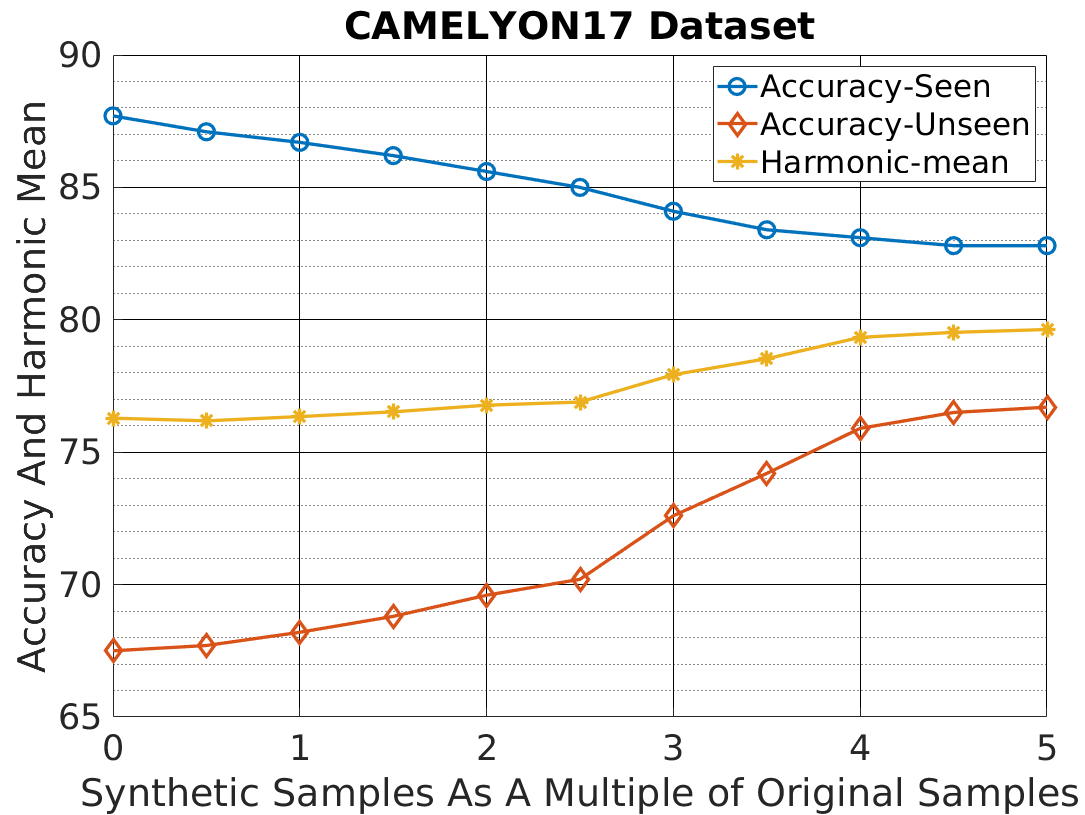}  \\
 (a) & (b) \\
\end{tabular}
\caption{Value of accuracy and H when adding synthetic samples to the dataset: (a) NIH dataset; (b) CAMELYON17 dataset.}
\label{fig:PlotX}
\end{figure}

%
Figure~\ref{fig:PlotX} shows, for the NIH Chest Xray and CAMELYON17 dataset, the effect of adding synthetic samples on $Acc_S, Acc_U$ as a function of dataset augmentation factor. 
Increasing synthesized examples increases $Acc_U$ at a high rate while reducing $Acc_S$, although at a lower rate. T
Synthetic samples improve discriminative power of classifiers and reduce bias towards Seen classes. 

%

\section{Conclusion}

We propose a GZSL approach for medical images without relying on class attribute vectors.  
Our novel method can accurately synthesize feature vectors of unseen classes by employing self supervised learning at different stages such as anchor vector selection, and training the feature generator. Using self supervision allows us to bridge the semantic gap between Seen and Unseen classes. The distribution of synthetic features generated by our method are close to the actual distribution, while removing the self-supervised term results in unrealistic distributions.
Experimental results show our method outperforms other GZSL approaches in literature.  


{\small
\bibliographystyle{ieee_fullname}
\bibliography{ICCV2021_GZSL}
}

\end{document}